\documentclass[conference]{IEEEtran}
\IEEEoverridecommandlockouts
\usepackage{cite}

\usepackage{amsmath,amssymb,amsfonts}
\usepackage[linesnumbered,ruled,vlined]{algorithm2e}
\usepackage[dvips]{graphicx}
\usepackage{textcomp}
\usepackage{xcolor}
\usepackage[T1]{fontenc}
\usepackage{colortbl}
\usepackage{subfigure}
\usepackage{verbatim,float,caption}
\usepackage[left=0.59in,right=0.59in,bottom=0.95in,top=0.8in]{geometry}

\def\BibTeX{{\rm B\kern-.05em{\sc i\kern-.025em b}\kern-.08em
    T\kern-.1667em\lower.7ex\hbox{E}\kern-.125emX}}

\begin{document}
	
		\title{\huge Smart Resource Allocation Model via Artificial Intelligence in Software Defined 6G Networks
		\author{ Ali Nouruzi$^{*}$, Atefeh Rezaei$^{\dag}$, Ata Khalili\IEEEauthorrefmark{3}, Nader Mokari$^{*}$, Mohammad Reza Javan$^{\S}$, Eduard A. Jorswieck$^{\$}$, and \\Halim Yanikomeroglu$^{\pounds}$\\$^{*}$Department  of Electrical and Computer Engineering, Tarbiat Modares University, Tehran,~Iran\\~$^{\dag}$Technical University of Berlin, Germany;\\$^{\ddag}$Friedrich-Alexander-University Erlangen-Nurnberg, Germany\\~$^{\S}$Department of Electrical and Robotic Engineering, Shahrood University of Technology, Shahrood, Iran\\~$^{\$}$Institute of Communications Technology, Technische Universität Braunschweig, Braunschweig, Germany\\~$^{\pounds}$Non-Terrestrial Networks (NTN) Lab, Department of Systems and Computer Engineering, Carleton University, Ottawa}
				\thanks{The work of Eduard Jorswieck was supported in part by the Federal Ministry of Education and Research (BMBF, Germany) in the program of "Souverän. Digital. Vernetzt." joint project 6G-RIC, project identification number: 16KISK020K and 16KISK031.}}
	\maketitle

\begin{abstract}
In this paper, we design a new flexible smart software-defined radio access network (Soft-RAN) architecture with traffic awareness for sixth generation (6G) wireless networks. 
In particular, we consider a hierarchical resource allocation model for the proposed smart soft-RAN model where the software-defined network (SDN) controller is the first and foremost layer of the framework. This unit dynamically monitors the network to select a network operation type on the basis of distributed or centralized resource allocation procedures to intelligently perform decision-making. 
In this paper, our aim is to make the network more scalable and more flexible in terms of conflicting performance indicators such as achievable data rate, overhead, and complexity indicators. To this end, we introduce a new metric, i.e, throughput-overhead-complexity (TOC), for the proposed machine learning-based algorithm, which supports a trade-off between these performance indicators. 
In particular, the decision making based on TOC is solved via  deep reinforcement learning (DRL) which determines an appropriate resource allocation policy.
 Furthermore, for the selected algorithm, we employ the soft actor-critic (SAC) method which is more accurate, scalable, and robust than other learning methods.~Simulation results demonstrate that the proposed smart network achieves better performance in terms of TOC  compared to fixed centralized or distributed resource management schemes that lack dynamism. Moreover, our proposed algorithm outperforms  conventional learning methods employed in recent state-of-the-art network designs.
	\end{abstract}
\section{Introduction }
The evolution from fifth-generation (5G) to sixth-generation (6G) wireless networks is proceeding along two main lines, which consist of developing network architecture and communications technologies. As 5G and beyond networks continue to evolve, it will be possible for network architectures and communication technologies to be dynamically adapted according to the changing demands of the network. In such a flexible architecture, a huge amount of signaling and computational resources are needed to manage the network resources efficiently. 

Due to 
discrepancies between the mathematical tractability
and the exponentially greater complexity of
wireless networking, conventional resource allocation approaches are unfortunately ineffective and may not be able to fulfill 
the precise quality of service (QoS) requirements
of emerging applications.
Furthermore, 6G networks require to provide performance based on the different conditions, such as traffic changes or the variety  of services requested by users.
To tackle these issues, artificial intelligence (AI) has been identified as a promising solution for automatic and autonomous network
management. Adopting intelligent resource allocation
for wireless networks not only has the potential
to replace the manual intermediation needed for current network management
tasks, but also presents novel 
optimization possibilities to ameliorate performance gains online in real-time.~Since future dense wireless networks will involve high complexity algorithms, machine learning (ML) has emerged as a key enabler to manage high complexity for real-time
implementation. In this regard, reinforcement learning (RL) as a type of ML has been employed to learn from an environment by trial and error, which promotes improvements over time. Also,  deep reinforcement learning (DRL)  has been investigated for comprehensive inputs as well as more accurate results compared to  RL algorithms \cite{mag}. Furthermore, to enable a real-time scheduler for stochastic environments, it has been shown that learning via multiple agents can solve complicated stochastic optimization problems \cite{SAC_mec}.
%~Although DRL methods have been applied to several resource allocation problems, they have two major challenges: high sample complexity and sensitivity to hyper-parameters\cite{SAC}.  
Besides, the soft actor-critic (SAC) algorithm with a maximum entropy objective can
be leveraged to provide more accurate and stable solutions for  scalable networks in dynamic environments compare to the baselines.

%\subsection{Resource Allocation for Wireless Communication}
%Let us start by introducing the candidate transmission strategy and the resource allocation frameworks in  conventional network architectures.~
ML algorithms have already been exploited for resource multiple allocation in a wireless communication system to obtain an efficient solution from instantaneous actual channel data. Also, thanks to the capability of ML solutions in adaptation to the environment, they lead to a higher performance
in practice.~In this regards, there are a plethora of works that considered resource allocation based on ML algorithms [2]-[6]. In \cite{SAC_mec}, a multi-access edge computing technique was considered that reduced the core network congestion. More specifically, cooperative computation offloading policy was designed for mobile edge computing (MEC) technology using the SAC method for both the centralized and distributed offloading. The authors in \cite{Mu} proposed a matching-based distributed structure  to improve the latency and proficiency of a network.
In \cite{zhang2019deep}, a DRL algorithm for vehicular-to-everything (V2X) communication was proposed that determined resource block allocation and performed power control. The DRL algorithm has the ability to select the transmission mode (i.e., vehicle-to-infrastructure (V2I) or vehicle-to-vehicle (V2V) communications.
The authors in \cite{Xia1} performed functional splits
of control and data planes between the cloud and edge nodes in C-RAN while taking the fronthaul delay into account.
% In \cite{Kang1}, the cloud-edge allocation of the control function of rate selection and the data plane function of decoding on uplink communication in the Fog-aided network architectures is proposed.
The authors in \cite{atefeh} considered a semi-centralized framework for the resource allocation problem by using matching theory and a successive convex approximation (SCA) approach.
%Similarly, the authors in \cite{multiagent}  proposed a semi-centralized resource allocation scheme to maximize the weighted matching (MWM) problem  for  integrated access and backhaul (IAB) networks.
The authors in \cite{jointcendes} compared both centralized and distributed algorithms regarding the base band unit (BBU)  location placement problem in C-RAN, where their proposed solution is based on a distributed heuristic algorithm.
Moreover, in \cite{valaee1}, the authors compared the energy efficiency of their proposed distributed and centralized user association algorithms by sequentially minimizing the power consumption of the heterogeneous network.
Motivated by the above discussion, in this paper we propose an intelligent approach for determining an effective resource allocation policy.
We consider a network that can switch between  centralized and distributed operations for  on demand resource management. 
To the best of our knowledge, a smart network architecture configuration such as this has not been studied yet. 
%In  Table \eqref{Tab:REF} we summarized the literature and highlight the targets of this papers. 
Previous works have not considered the changing environmental conditions of networks, and thus they are not appropriate for changes in real scenarios. Besides, the complexity, overhead, and achievable data rate performance metrics are not considered jointly in the other existing works in the literature while it is important to consider them jointly in selecting the proper resource management.
%The capacity of networks to learn autonomously and change their architectures to boost performance will be a critical enabler of next-generation intelligent wireless networks. 
The main contributions of this paper are summarized as follows:
\begin{itemize}
	\item In contrast to the conventional approaches that employ
	a model for the resource allocation based on the analytical models, an
	intelligent approach based on the learning methods is exploited for solving the  software-defined network (SDN) decision and resource allocation problems in a centralized or distributed manner.~This approach allows a network to adapt to the environment and perform more effectively in real-world scenarios. In particular, we introduce a novel algorithm for the SDN controller that chooses the best resource allocation policy based on the DRL method. In this model, the centralized or distributed modes are selected on the basis of a throughput metric that considers overhead, complexity, and the total data rate. Hence, our proposed smart algorithm decides whether a BBU or each re-configurable radio systems (RRS) is responsible for the resource allocation policy. 
	\item In the centralized scheme, we consider the single-agent SAC algorithm at the centralized unit that designates the appropriate actions based on the collected information. Additionally, in the distributed scheme, we consider that there is no information exchange between agents, and that the BSs (as RRSs) locally perform resource allocation tasks based on a multi-agent actor-critic algorithm. This algorithm can be applied to  more complex environments and  provides stable solutions for the network compared to the other learning methods.  
	\item Numerical results indicate a performance gain by employing the SAC methods relative to other ML based (i.e., DDPG and DQN approaches). Furthermore, results demonstrate performance gaps between fixed centralized or fixed distributed schemes with the proposed smart algorithm.
	
\end{itemize}

\section{System Model and Problem Statement}
\label{sec_systemmodel}
As shown in Fig. \ref{system_model}, we consider that the RAN network consists of two major units which we refer to as the access network outer (AN-O) and the access network inner (AN-I). The AN-O is the cloud RAN network, and the AN-I includes $B$ RRSs as local units. The RRSs provide signaling and data coverage for the users.  As a result of considering an intelligent network, it is necessary to obtain some efficient feedback about the network status to operate in a real-time.
Furthermore, it is assumed that the information can be fed back from the BSs to the AN-O through the control links. 
 Also, we assume that all RRSs and users are equipped with single antenna. The AN-O consists of a BBU pool which performs centralized baseband processing, and a centralized SDN controller, which controls the network operation by programming the network's element functionalities properly.
In particular, we consider that the SDN controller saves a stream of data rates that was achieved from the previous time slots, and it stores the flows including the overheads and complexities in its buffer.
By monitoring the network, the SDN controller determines whether the network would operate in a centralized or distributed manner.

\subsection{Decision-Making Optimization Problem}
We consider a comprehensive framework in which the SDN controller intelligently switches between  centralized and distributed network operations by considering the total data rate and the amount of data exchanged in terms of overhead and complexity.
We define two binary decision making parameters for the centralized and distributed scenarios as $x_{\text{Cnt}}^{(t)}$ and $x_{\text{Dst}}^{(t)}$, respectively, where $\boldsymbol{x}^{(t)}=\{x_{\text{Cnt}}^{(t)},x_{\text{Dst}}^{(t)}\}$. To make a trade-off between data rate, overhead, and complexity, we introduce a new metric term as TOC to perform mode selection.~Consequently, the decision-making structure for the SDN controller is shown at the top of Fig. \ref{system_model}.
 In this framework, we consider that in each  time slot $t$, only one operation scheme can be selected as follows: 
\begin{align}
x^{\left( t\right) }_{\text{Dst}}+x^{\left( t\right) }_{\text{Cst}}=1, \forall t, x^{\left( t\right) }_{\text{Dst}}, x^{\left( t\right) }_{\text{Cst}} \in \{0,1\}. 
\end{align}
 In the following, we introduce and model the three components of TOC:

\begin{figure*}
	\centering
	\includegraphics[width=0.4\textwidth]{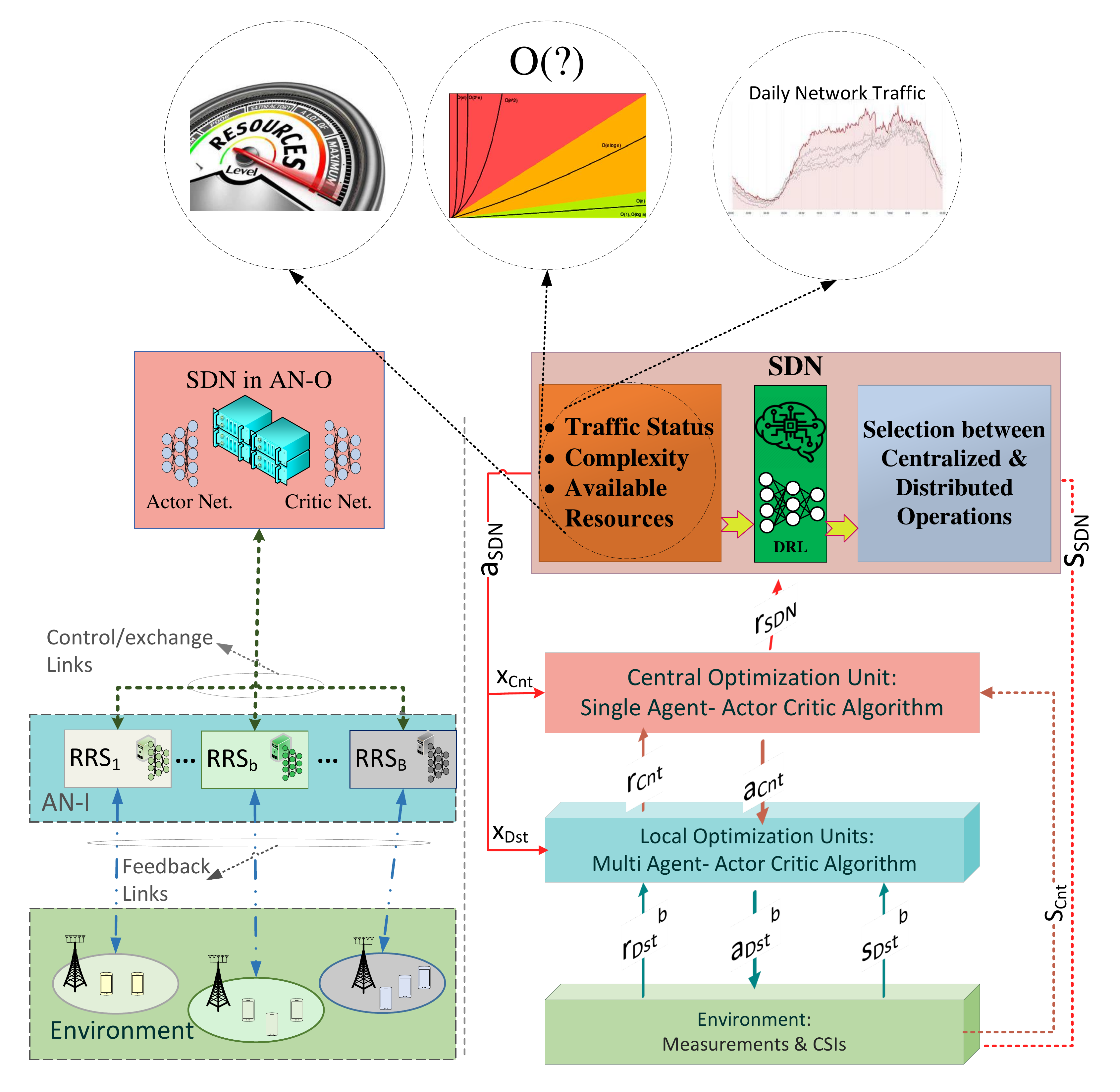}
	\caption{System model and hierarchical flowchart  of the smart soft-RAN.}
	\label{system_model}
\end{figure*}
\subsubsection{Overhead} 
As a critical key performance indicator, we  address the overhead in the telecommunication networks. The overhead is a function of the number of information bits needed to feed back  the channel status, subcarrier indicators, and the transmission power of a specific user over a subcarrier.
Also, the total number of  RRSs, users, and subcarriers in each RRS and in each time slot can affect the  overhead.
In the centralized mode, the resource allocation operation is performed at the BBU, and thus the information needs to be transmitted from the RRSs to the centralized unit and it is denoted by $\tau^{t}_{\text{Cnt}}$ at time slot $t$.  In the distributed mode, by contrast, all operations related to  resource allocation are performed independently by the RRSs without any data exchange  and it is denoted by $\tau^{b,t}_{\text{Dst}}$  for each of RRSs. 
Accordingly, $\tau^t_{\text{Cnt}}$ and $\tau^{b,t}_{\text{Dst}}$ are calculated as follows:
\begin{align}
&\tau^{b,t}_{\text{Dst}} = \left( \beta_{\text{Power}} + \beta_{\text{CSI}} + \beta_{\text{Subcarriers}} \right) \cdot |\mathcal{U}^{t}_b|\cdot|\mathcal{K}^{t}_b|,
\\
&\tau^{t}_{\text{Cnt}} = \sum_{b\in \mathcal{B}}\tau^{b,t}_{\text{Dst}},
\end{align}
where $\beta_{\text{Power}}$, $\beta_{\text{Subcarriers}}$, and $\beta_{\text{CSI}}$ represent the number of required bits to describe the power allocation variable matrix, subcarrier allocation  variable matrix, and the channel state information (CSI) matrix.
In addition, the size of the set of users that are associated with RRS $b$ is denoted by $|\mathcal{U}^t_b|$. Moreover, $|\mathcal{K}^t_b|$ represents the size of the set of subcarrier for RRS $b$. Ultimately, the set of total users  and subcarriers are denoted by $\mathcal{U}$ and $\mathcal{K}$, respectively.

It is remarkable that for the decision making procedure in addition to the achieved  data rates of the centralized scheme, the distributed data rates in a certain number of previous time slots are fed back and stored in the  SDN controller's memory. Hence, the overhead of decision making procedure originates from exchanging the data rates of remote radio heads (RRH) and SDN controller which is tolerable.
Also, to reduce the overhead of signaling  in the distributed operation, we consider that the decisions of each specific RRS regarding its allocated powers, subcarriers and the estimated channels are not exchanged with the other distributed RRSs.
\subsubsection{Achievable Data Rate}
Let binary variable $\rho^{t,b,u,k}\in\left\lbrace 0,1\right\rbrace $  denote that subcarrier $k$ in RRS $b$ for user $u$ at time slot $t$ and its  transmission power is considered as $p^{t,b,u,k}\ge0$.

 Consequently, for the centralized resource allocation scheme, $\mathbf{P}^t_{\text{Cnt}}=\left[ p^{t,b,u,k}_{\text{Cnt}}\right] $ and  $\boldsymbol{\rho}^t_{\text{Cnt}}=\left[ \rho^{t,b,u,k}_{\text{Cnt}} \right] $ and for the distributed scheme $\mathbf{P}^{t}_{\text{Dst},b}=\left[ p^{t,b,u,k}_{\text{Dst}}\right] $ and  $\boldsymbol{\rho}^{t}_{\text{Dst},b}=\left[\rho^{t,b,u,k}_{\text{Dst}}\right] $.
In addition, the channel gain between user $u$ and RRS $b$ on subcarrier $k$ at time slot $t$ is denoted by
 $h^{t,b,u,k}=h_\text{Large}^{t,b,u,k}h_\text{Small}^{t,b,u,k}$ 
which includes small scale fading and large scale fading effects. 
Accordingly, the total achieved data rate by
the centralized method is given by.
\begin{align}
	r^t_{\text{Cnt}} = \sum_{b\in\mathcal{B}} \sum_{k \in \mathcal{K}}\sum_{u \in \mathcal{U}} \left( 1+\frac{h^{t,b,u,k} p^{t,b,u,k}_{\text{Cnt}}\rho^{t,b,u,k}_{\text{Cnt}}}{\sigma^2+I^{t,b,u,k}_{\text{Cnt}}} \right), \forall t,
\end{align}
where $\sigma^2$ is the noise variance power of AWGN. In addition,  $I^{t,b,u,k}_{\text{Cnt}}$ is the inter-cell interference on subcarrier $k$ in RRS $b$  for user $u$ and it is calculated as follows:
\begin{align}
	I^{t,u,k}_{\text{Cnt}}=\sum_{b'\ne b \in \mathcal{B}}\sum_{u' \ne u \in \mathcal{U}} h^{t,b',u,k} p^{t,b',u',k}_{\text{Cnt}}\rho^{t,b',u',k}_{\text{Cnt}}.
\end{align}
In a similar manner, the total achieved data rate by adopting the distributed scheme is denoted by $r^t_{\text{Dst}}$ and is given by.
\begin{align}
	r^t_{\text{Dst}} = \sum_{b\in\mathcal{B}} \sum_{k\in\mathcal{K}_b}\sum_{u \in \mathcal{U}_b} \left( 1+\frac{h^{t,b,u,k} p^{t,b,u,k}_{\text{Dst}}\rho^{t,b,u,k}_{\text{Dst}}}{\sigma^2+I^{t,u,k}_{\text{Dst}}} \right), \forall t,
\end{align} 
where $ I^{t,u,k}_{Dst}
 $ is the inter-cell interference for user $ u $ at time slot $ t $.
 It can be seen that 
 in the distributed manner the interference only depends on the user location, i.e., the large scale fading as each RRS calculates the large scale fading of the other RRSs to its users based on the locations. Also, in the worst-case scenario, we introduce an upper bound for the interference function 
  which is based on the maximum transmit power that each RRS can allocate. Thus, the inter-cell interference is computed as follows:
\begin{align}
I^{t,u,k}_{\text{Dst}} = \sum_{b'\ne b \in\mathcal{B}}\sum_{u'\ne u \in\mathcal{U}} h_{\text{Large}}^{t,b',u}P^{b'}_{\text{Equal}},
\end{align}
in which $ P^{b'}_{\text{Equal}}= \dfrac{p_{\text{max}}^{b'}}{|\mathcal{U}^t_{b'}||\mathcal{K}^t_{b'}|} $.
\subsubsection{Computational Complexity} 
The computational complexity of  DRL-based methods depends on the number of neurons,  layers, the state size, the action space, the number of episodes ($E$), and the number of samples of random batch ($M$) \cite{ATA}. 
Accordingly, by considering a DRL network  with $|N|$ hidden layers, the total computational complexity for the centralized method, is calculated as follows:
\begin{align}
\Gamma_{\text{Cnt}}^{t} = O \bigg(E \cdot M \cdot \big(l^{\text{Input}}_{\text{Cnt},t}l_1+\sum^{N-2}_{n=1} l_{n}l_{n+1}+l_{N-1}l^{\text{Output}}_{\text{Cnt},t} \big)\bigg) ,
\end{align}
where $l_{n}$ is the number of neurons in the layer $n$ of the deployed neural network, and $l^\text{Input}_{\text{Cnt},t}$ is the number of input layer  in the DRL network and it is equal to the state size which is given by $l^\text{Input}_{\text{Cnt},t}=|\mathcal{U}^t|\cdot|\mathcal{K}^t|\cdot|\mathcal{B}|$. In addition,  $l^\text{Output}_{\text{Cnt},t}$ denotes the number of output layers for the centralized method and it is calculated by $l^\text{Output}_{\text{Cnt},t}=|\boldsymbol{P}^{t}_{\text{Cnt}}|+|\boldsymbol{\rho}^{t}_{\text{Cnt}}|$.
In a similar manner, the computational complexity for the distributed method in RRS $b$ is given by .
\begin{align}
\Gamma^{b,t}_{\text{Dst}} =O\bigg(  E \cdot M \cdot \big(l^{\text{Input}}_{\text{Dst},b,t}l_1+\sum^{N-2}_{n=1} l_{n}l_{n+1}+l_{N-1}l^{\text{Output}}_{\text{Dst},b,t} \big)\bigg) , 
\end{align}
where $l^{\text{Input}}_{\text{Dst,b,t}}$ denotes the number of neurons in the input layer of DRL in the RRS $b$ and it is given by $l^{\text{Input}}_{\text{Dst,b,t}}=|\mathcal{U}^t_b|\times|\mathcal{K}^t_b|$.
Moreover,  $l^{\text{Output}}_{\text{Dst,b,t}}$ is the number of neurons in the output layer for RRS $b$ and it is given by $l^{\text{Output}}_{\text{Dst,b,t}}= |\boldsymbol{P}^{t}_{\text{Dst},b}|+|\boldsymbol{\rho}^{t}_{\text{Dst},b}|$.
%\textcolor{blue}{
\subsubsection{Main Problem}
%In this framework, we consider that in each transmission time slot $ t $ only one operation scheme can be selected. 
%With the aim of maximizing TOC, the following optimization problem is introduced:
%
%in which $\alpha$ and $\beta$ are the coefficient factors related to the overhead and complexity. In addition, constraint \eqref{pow} insure that the total allocated power in each of BSs is lower than maximum transmission power.
%%}

\begin{figure}
	\centering
	\includegraphics[width=1\linewidth]{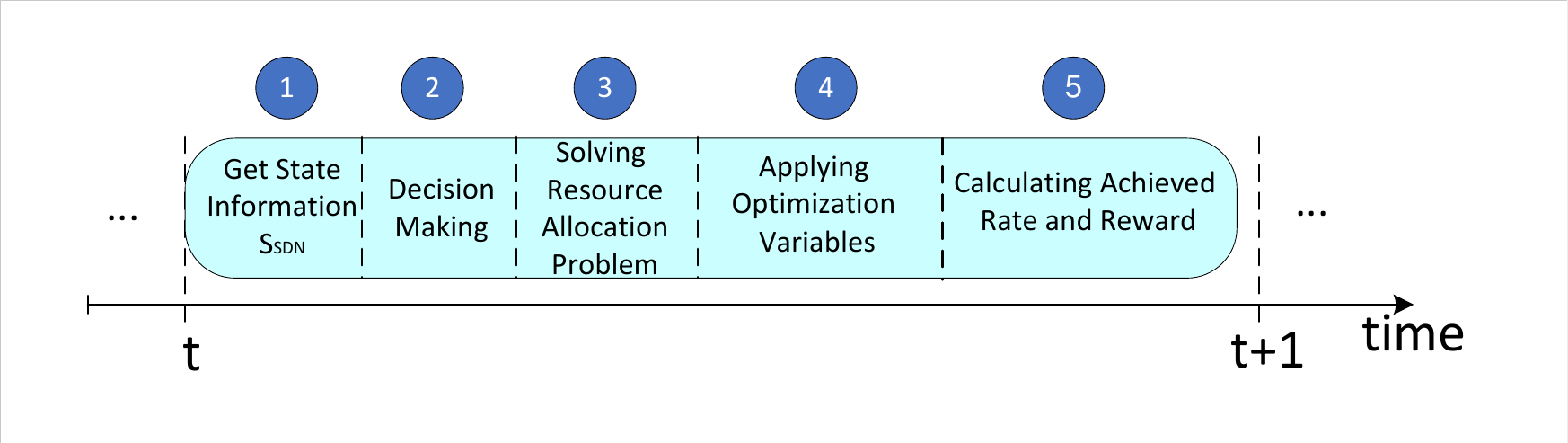}
	\caption{The details of method selection process in the time.}
	\label{fig:time-frame}
\end{figure}

The SDN controller stores data rates, overhead, and complexities of the $D$ previous time slots  as follows: $ \textbf{r}_{Cnt}=\{r_{Cnt}^{t-D},...,r_{Cnt}^{t-1}\} $, $ \textbf{r}_{Dst}=\{r_{Dst}^{t-D},...,r_{Dst}^{t-1}\} $,
$ \boldsymbol{\tau}_{Cnt}=\{\tau_{Cnt}^{t-D},...,\tau_{Cnt}^{t-1}\} $,
$ \boldsymbol{\tau}^{b}_{Dst}=\{\tau_{Dst}^{t-D},...,\tau_{Dst}^{t-1}\} $,
$ \boldsymbol{\Gamma}_{Cnt}=\{\Gamma_{Cnt}^{t-D},...,\Gamma_{Cnt}^{t-1}\} $,
$ \boldsymbol{\Gamma}_{Dst}^{b}=\{\Gamma_{Dst}^{t-D},...,\Gamma_{Dst}^{t-1}\} $. Also, based on the previous stored data, the following TOC metrics are considered:

\begin{align}
\label{TOC1}
& TOC^{t}_{\text{Cnt}}=r^{t}_{\text{Cnt}} -\beta\tau^{t}_{\text{Cnt}}-\alpha \Gamma^{t}_{\text{Cnt}},
\\
\label{TOC2}
& TOC^{t}_{\text{Dst}}=
r^{t}_{\text{Dst}} -\beta\max\{\tau^{b,t}_{\text{Dst}}\}-\alpha \max\{ \Gamma^{b,t}_{\text{Dst}}\},   
\end{align}
where $\alpha$ and $\beta$ are the coefficient factors related to the overhead and computational complexity, which is affected by the number of information updates to solve the allocation problem resources.
To have a better representation of the smart process of choosing solution method over the time, we provide  the time-based scheme illustrated in Fig. \ref{fig:time-frame}.

% Consequently, the proposed ML algorithm consider $ \boldsymbol{\xi}^t_\text{Cnt}=\{TOC_\text{Cnt}^{t-D-1},...,TOC_\text{Cnt}^{t-1}\} $ and $ \boldsymbol{\xi}^t_\text{Dst}=\{TOC_\text{Dst}^{t-D-1},...,TOC_\text{Dst}^{t-1}\} $ as the  rewards to make its decision for the $ t^{th} $ time slot.
%\begin{algorithm}
%	\caption{Smart Resource Allocation Policy Selection}
%	Initialize the weights for actor and critic networks as $\theta$ and $\omega$
%	\\
%	Initialize the replay buffer $\mathcal{D}$
%	\\
%	Set the memory of selected methods $\mathcal{M}\leftarrow$[]
%	\\
%	\For{each time slot $t$}{
%		Considering the network state $S^{(t)}_{\text{SDN}}$  and select a appropiriate policy for resource allcation:
%		\\
%		\If{ $x_{\text{Cnt}}=1$ }{Calculate the TOC from \eqref{TOC}}
%		\Else{Calculate the TOC from \eqref{TOC2}
%		}
%	}
%\end{algorithm}
\begin{algorithm}[h!]
	\small
	\renewcommand{\arraystretch}{0.4}
\caption{Smart Resource Allocation Policy Selection Based on SAC}
	\label{SAC2}
	%\textbf{Hyperparameters:} Step sizes $\lambda_{\pi}, \lambda_{Q}, \lambda_{\varphi}$, target entropy $\mathfrak{e}$, exponentially moving average coefficient $\tau$ 
	\textbf{Input:} Initial Q value function parameters $\phi$
	\\\textbf{Input:} Initial policy parameters $\theta$
	\\\textbf{Set:} The memory of transaction $\mathcal{B}=\emptyset;$ 
	\\\For{each iteration}{
		\For{each environment step}{
		In state $S^{(t)}_{\text{SDN}}$, based on the policy $\pi$ with parameters $\theta$, select action $a^{(t)}_{\text{SDN}}$:
		\\
			$\nonumber 
			a^{(t)}_{\text{SDN}}\sim\pi_{\theta}\left(\cdot|S^{(t)}_{\text{SDN}}\right)$
			\\
			\If{$x^{\left( t\right) }_{\text{Cnt}}=1$}{			Based on \eqref{TOC1}, calculate the reward.}
			\Else{Based on \eqref{TOC2}, calculate the reward
			network state changes: $S^{(t)}_{\text{SDN}}\leftarrow S^{(t+1)}_{\text{SDN}}$
			\\
			$\mathcal{B}\leftarrow\mathcal{B}\cup\left\{S^{(t)}_{\text{SDN}},a^{(t)}_{\text{SDN}},r\left(S^{(t)}_{\text{SDN}},a^{(t)}_{\text{SDN}}\right),S^{(t+1)}_{\text{SDN}}\right\}$
		}
	}
		\For{each gradient step}{
			$\nonumber $ 
			Update parameters   
			$\theta$ and $\phi$ based on the  Adam\cite{adam} optimizer.	      
		}
	}	                      
\end{algorithm}
\subsection{Proposed Solution for the SDN}
\begin{figure*}
	\centering
	\includegraphics[ width=0.9\textwidth]{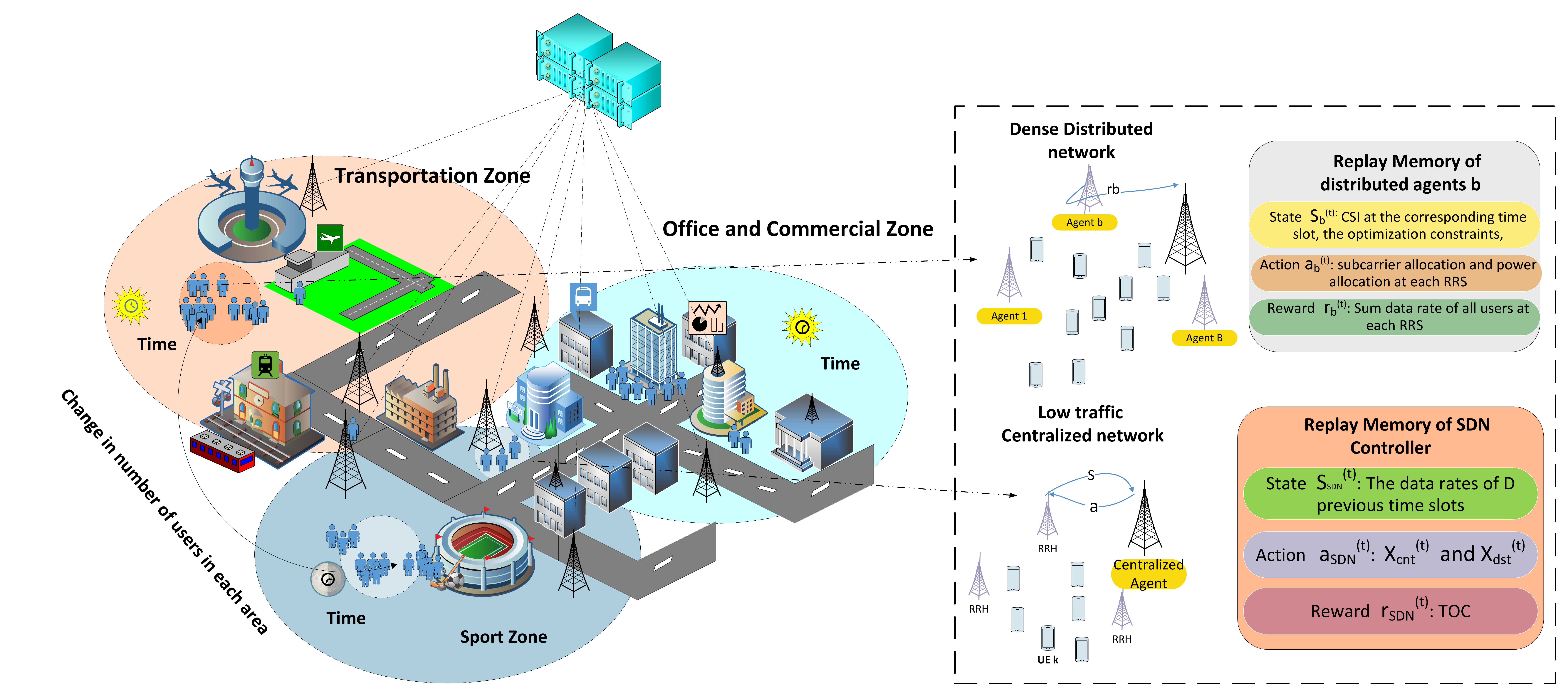}
	\caption{Network architecture based on the actor critic framework.}
	\label{distcent}
\end{figure*}

 We employ the DRL framework for the decision-making problem at the SDN controller in which the states, actions, and rewards are described in Fig. \ref{distcent}. In the SDN algorithm, based on the given state $ S_{\text{SDN}}^{(t)} $ in each time slot, the SDN  selects an appropriate action $ a_{\text{SDN}}^{(t)} $. 
As you can see in Fig. \ref{system_model}, we consider a  memory of  network traffic and users' channel state information as $ S_{\text{SDN}}^{(t)} $.
By taking the action at each time slot, the agent gets a reward $ r_{\text{SDN}}^{(t)} $  that is defined on the basis of the objective function in the main resource allocation problem. After observing $ S_{\text{SDN}}^{(t+1)} $, the SDN stores $S_{\text{SDN}}^{(t)}, a_{\text{SDN}}^{(t)}, r_{\text{SDN}}^{(t)}  $, and $ S_{\text{SDN}}^{(t+1)} $. Then, it employs a  stochastic optimization method like Adam\cite{adam} to minimize the loss function and update its decision. 
The pseudo code of the proposed solution is provided in Algorithm. \ref{SAC2}.
\begin{figure*}[t]
	\centering
	\centerline{\includegraphics[width=0.5\textwidth]{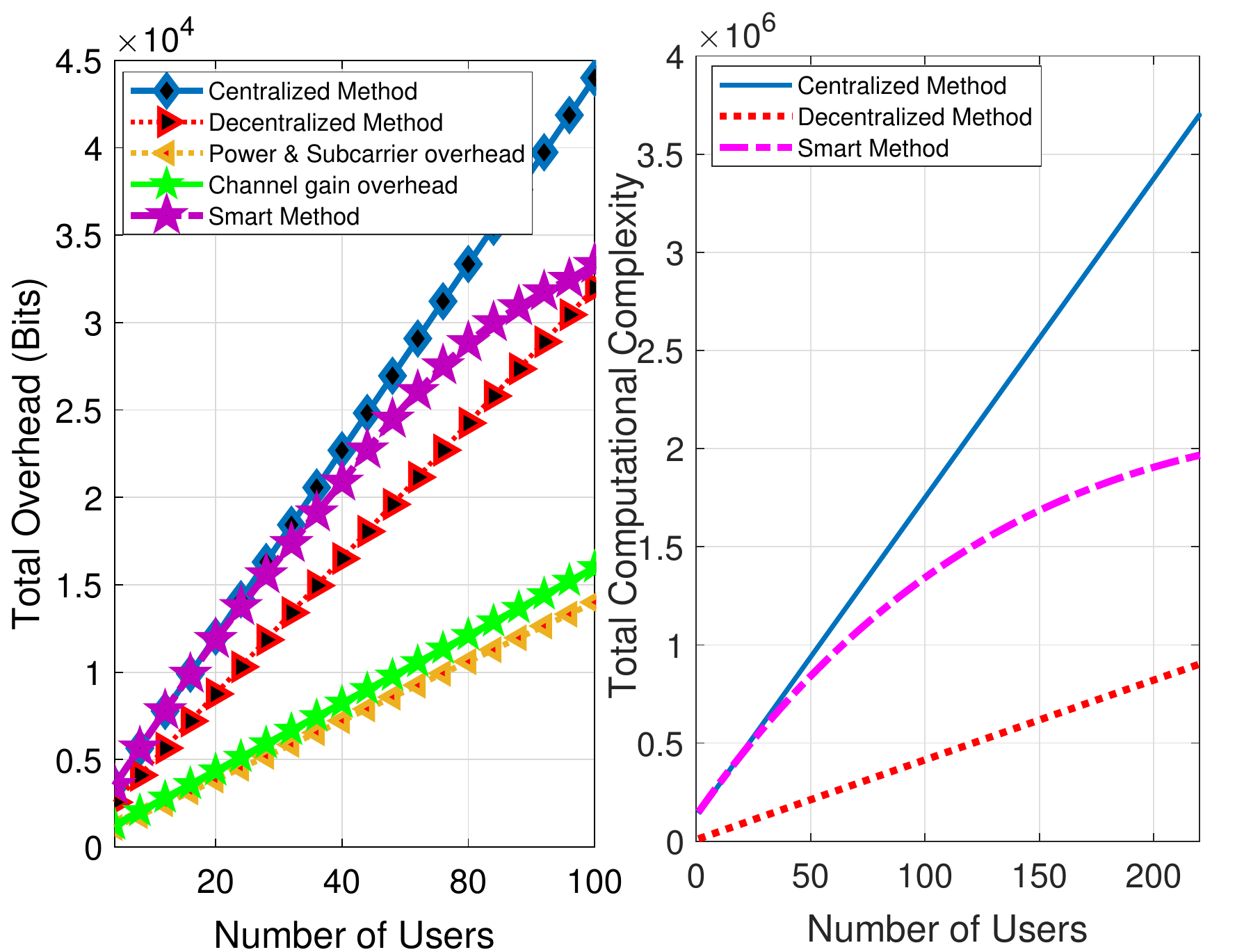}}
	\caption{Total overhead and complexity relative to the number of users.}
	\label{overhead}
\end{figure*}
\subsection{Resource Allocation Problem}
After the decision-making stage, the resource allocation problem (RAP) is solved using one of the two different schemes in terms of centralized or distributed.~In particular, we formulate the resource allocation problem to maximize the throughput of the network while taking into account the subcarrier allocation and power control based on the downlink OFDMA.
To solve the problem, we propose a SAC based resource allocation for the centralized and distributed scenarios.~In what follows, we explain how the resource allocation problem can be solved.\\
%\subsection{Proposed Solution for the SDN} Here, we provide an efficient algorithm for deciding between the centralized and distributed schemes. The DRL framework for the decision-making problem at the SDN is employed whose states, actions, and rewards are described in Fig. \ref{distcent}. In the SDN algorithm, based on the given state $ S_{\text{SDN}}^{(t)} $ in each time slot, the SDN  selects an action $ a_{\text{SDN}}^{(t)} $. By taking the action at each time slot, the agent gets a reward $ r_{\text{SDN}}^{(t)} $  that is defined on the basis of the objective function in $ \mathcal{P}_{\text{SDN}} $. After observing $ S_{\text{SDN}}^{(t+1)} $, the SDN stores $S_{\text{SDN}}^{(t)}, a_{\text{SDN}}^{(t)}, r_{\text{SDN}}^{(t)}  $, and $ S_{\text{SDN}}^{(t+1)} $ and using Adam\cite{adam}, a method for efficient stochastic optimization, it minimizes the loss function and updates its decision.
%\subsection{Resource Allocation Problem} After the decision-making stage, the resource allocation problem (RAP) is solved using two different schemes.~In particular, we formulate the resource allocation problem to maximize the throughput of the network while taking into account the subcarrier allocation and power control based on the OFDMA. To solve the problem, we propose a soft actor-critic based resource allocation for the centralized and distributed scenarios.~In what follows, we explain how we solve the resource allocation problem.

\subsubsection{Centralized Scheme}   
In the case where $x^{\left(t \right) }_{\text{Cnt}}=1  $, the RAP is solved based on  a single agent SAC at the BBU pool for RRS $b$ as shown in Fig.~\ref{system_model}. Specifically, in this scheme, RRSs act as RRHs in which the radio frequency tasks and the resource allocation process are performed at the BBU pool. Each RRS collects related information and forwards it to the BBU pool. Based on the received information at time slot $t$, the agent chooses action $ a_{\text{Cnt}} $. Then, the actions are sent to the RRSs based on the resource allocated in the BBU pool unit. Moreover, the reward $ r_{\text{Cnt}} $ and new states $ S_{\text{Cnt}} $ are collected and forwarded to the BBU pool through fronthaul links.
\subsubsection{Decentralized Scheme}
In the distributed scheme $  \text{i.e., }   x^{\left( t\right)}_{\text{Dst}}=1$,  resource allocation is performed by the RRSs using an independent multi-agent actor critic method.
To reduce the overhead, we consider that RRS $b$  in the distributed mode cannot access  the others' information and just performs the resource allocation tasks by using its own information.
The  DQNs, are implemented at the RRS and each agent computes the power control and subcarrier allocation locally. As we can  see in Fig. \ref{distcent}, each RRS $ b $ at time slot $t$ observes state $ S_b^{(t)}  $ and takes the action $ a_{b}^{(t)} $, individually. Also, it receives the results of its own behavior as the reward $ r_b^{(t)}  $ without knowing the actions of the other agents. \label{Sec_Decision}
%The proposed algorithm for each actor-critic agent is detailed in Fig. \ref{DDRLF}.
% \begin{figure}
%	\centering
%	\includegraphics[width=8cm]{sac_a2.pdf}
%	\caption{Actor-critic algorithm in an agent.}
%	\label{DDRLF}
%\end{figure}

\section{Numerical Results and Discussion}
We consider a coverage area of radius $ 500$~m with four distributed BSs, each covering a radius of $100$~m.  The maximum transmit power of the BSs are set to $ 40$~dBm, and the total bandwidth is divided into $32$ orthogonal subcarriers.
The path-gain between a specific user and a BS for RF communication follows the Rayleigh distribution with distance-based path loss.  The noise power at each subcarrier is assumed $-174$ dBm/Hz.

%\textcolor{red}{
%\subsection{Overhead \& Complexity}
Fig. \ref{overhead} illustrates the overhead and complexity of the network based on the different policies (i.e., centralized, distributed, and smart). We assume the set $ \{16, \: 4, \: 4\} $ as the number of information bits to transmit channel status, subcarrier indicators, and the transmission power in the feedback process. As we can see, the network overhead and complexity for the centralized structure is too high, which leads to numerous difficulties in the network, such as the high delay and low reliability. The overhead and complexity in the distributed network are lower, which makes the distributed architecture preferable for real-time reliable networks. Fig. 4 also shows that the complexity and overhead of the centralized structure are greater in the ultra-dense networks, which makes the centralized policy inefficient for a dense network. This is due to the fact that in the centralized scheme, the complexity and overhead grow linearly due to resource management being performed over all the transmission nodes of the network with information being exchanged between all nodes. In contrast, the complexity and overhead of the distributed scheme are inferior.~This indicates that the complexity gap between the two schemes is much more sensible in dense networks with many users. On the other hand, the overhead and complexity of the smart model closely follow those of the centralized and distributed structures in low and ultra-dense networks, respectively. We remark that although for a single instance the smart decision selects one of the centralized/distributed approaches, as a result of the existence of different channels that are variable over the time, the overhead and complexity of the smart network on average would occur between two different strict centralized or distributed architectures.
\begin{figure} 
	\centering
	\centerline{\includegraphics[width=0.45\textwidth]{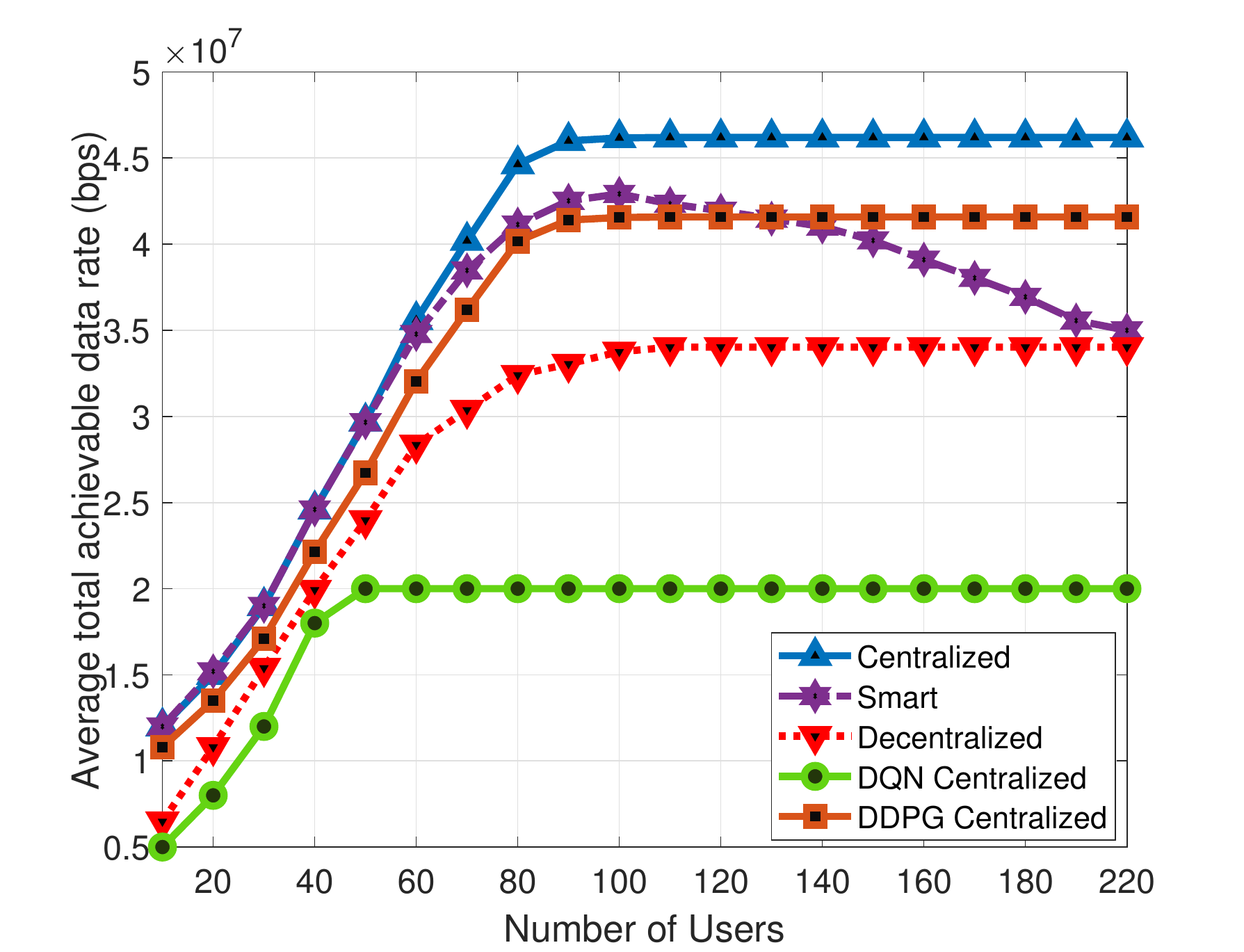}}
	\caption{Average total achievable data rate in bps relative to the number of users.}
	\label{pure}
\end{figure}

In Fig. \ref{pure}, we compare the achievable data rate for the different proposed architectures. As we can see, the achievable data rate of the centralized scheme outperforms the others, because in this scheme the network's knowledge is collected at the centralize unit, which raises the network awareness and efficiency. The smart solution achieves a data rate intermediate between the centralized and distributed schemes. In particular, by increasing the network's traffic, the smart scheme at first follows the centralized structure and then approaches the distributed case. Further, as Fig. \ref{pure} shows, the achievable data rate for a huge number of users in all schemes is saturated due to the limited resources. We remark that since there are different channel gains which are variable over time, the smart algorithm achieves the data rate between the centralized and distributed results.
We can also observe in Fig. 5 that the efficiency of the SAC based algorithm is greater than the other methods. This makes sense since the soft actor-critic can be employed for the continuous variables while the DQN needs some discretized variables, that results in some performance loss. 
%\subsection{New Performance Metric: Throughput Overhead Complexity (TOC)}
%In this section, we introduce a new metric for evaluating system performance as Throughput Overhead Complexity (TOC) which considers the effect of overhead and complexity in the network. 

\begin{figure}[t]
	\centering
	\centerline{\includegraphics[width=0.65\textwidth]{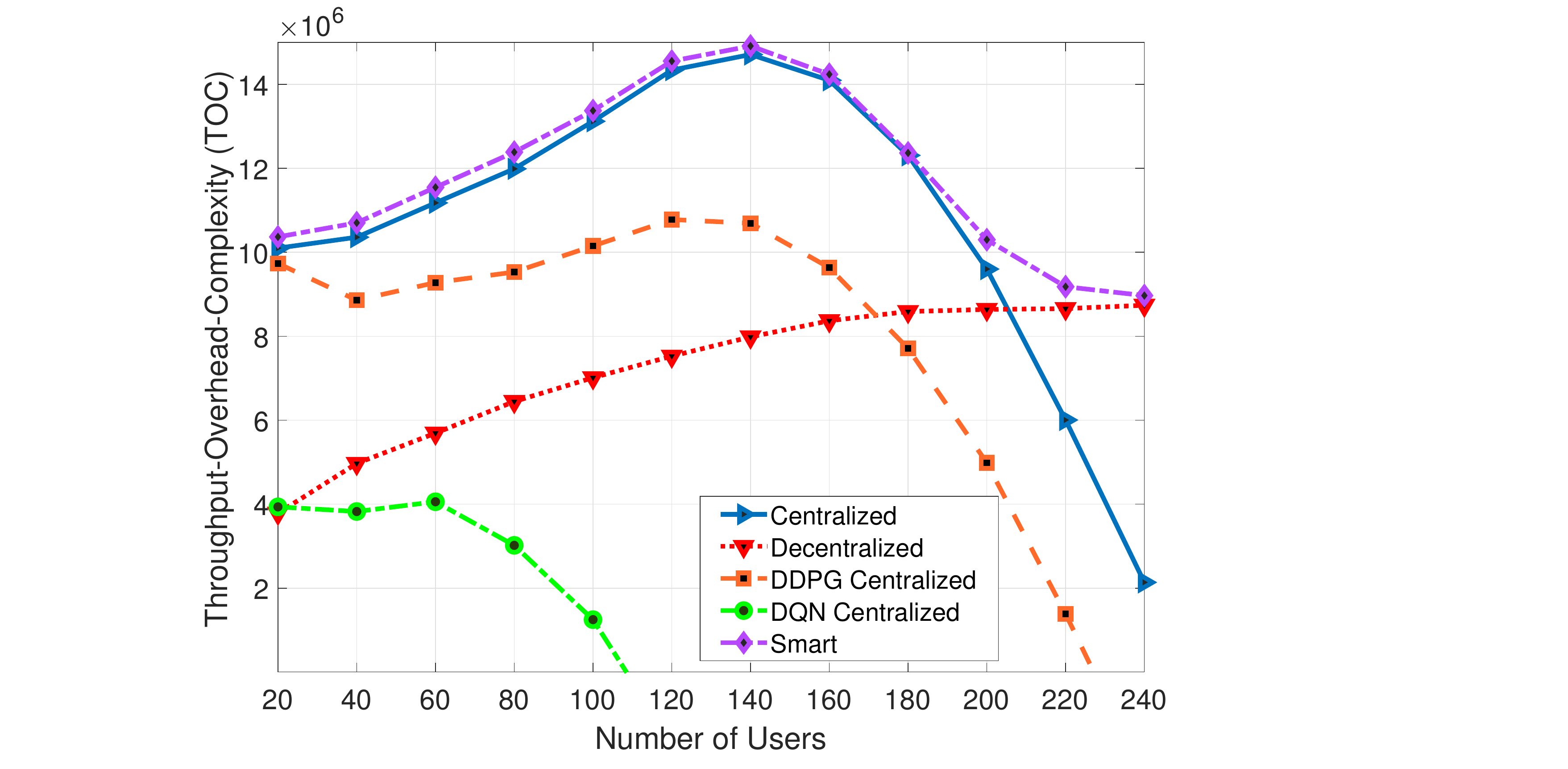}}
	\caption{TOC relative to the number of users.}
	\label{main_result}
\end{figure}

Fig. \ref{main_result} presents a comparison of Throughput Overhead Complexity (TOC) values for the centralized, distributed, and proposed smart network models. It is assumed that the number of users in the network varies depending on the level of traffic, from low to heavy. As we can see in Fig. \ref{main_result}, the centralized network shows a moderate performance increase in terms of TOC when the traffic status is low while it experiences a sharp decrease in a dense network. In the distributed network, there is a moderate increase of performance in terms of TOC by increasing the load of the network. At $ |\mathcal{K}|=200$, the throughput of the distributed structure exceeds that of the centralized one, and thus the decision-making algorithm here plays a vital role in the network.
Another interesting observation in this figure is that the performance gain in terms of the TOC through the learning-based approaches is better than that of the DDPG method.
This means that although the achieved data rate through the DDPG is higher than that of the ML solutions, the complexity of the DDPG in the centralized manner is very high, which deteriorates the performance gain in terms of TOC.
\section{Conclusion}
In this paper, we proposed a hierarchical network management model that adopts the best resource allocation policy in relation to changes in network status. The proposed intelligent model is density aware, which guarantees the network's performance on the basis of the DRL algorithm. We investigated three different scenarios (i.e., fixed centralized, fixed distributed, and dynamic) for different levels of network traffic.~Simulation results showed that the proposed algorithm not only performs better than conventional learning methods in terms of TOC, but it also outperforms both fixed centralized and distributed resource allocation policies.

\bibliographystyle{IEEEtran}
\bibliography{bib12}
\end{document}